# Why is Deep Random suitable for cryptology

Thibault de Valroger [(*)]

Abstract

We present a new form of randomness, called « Deep Randomness », generated in such a way that probability distribution of the output signal is made unknowledgeable for an observer. By limiting, thanks to Deep Randomness, the capacity of the opponent observer to perform Bayesian inference over public information to estimate private information, we can design protocols, beyond Shannon limit, enabling two legitimate partners, sharing originally no common private information, to exchange secret information with accuracy as close as desired from perfection, and knowledge as close as desired from zero by any unlimitedly powered opponent. We discuss the theoretical foundation of Deep Randomness, which lies on Prior Probability theory, introduced and developed by authors like Laplace, Cox, Carnap, Jefferys and Jaynes ; and we introduce computational method to generate such Deep Randomness.

### Introduction

Modern cryptography mostly relies on mathematical problems commonly trusted as very difficult to solve, such as large integer factorization or discrete logarithm, belonging to complexity theory. No certainty exist on the actual difficulty of those problems. Some other methods, rather based on information theory, have been developed since early 90's. Those methods relies on hypothesis about the opponent (such as « memory bounded » adversary [6]) or about the communication channel (such as « independent noisy channels » [5]); unfortunately, if their perfect secrecy have been proven under given hypothesis, none of those hypothesis are easy to ensure in practice. At last, some other methods based on physical theories like quantum indetermination [3] or chaos generation have been described and experimented, but they are complex to implement, and, again, relies on solid but not proven and still partly understood theories.

Considering this theoretically unsatisfying situation, we propose to explore a new path, where proven information theoretic security can be reached, without assuming any limitation about the opponent, who is supposed to have unlimited calculation and storage power, nor about the communication channel, that is supposed to be perfectly public, accessible and equivalent for any playing party (legitimate partners and opponents). In our model of security, the legitimate partners of the protocol are using Deep Random generation to generate their secret information, and the behavior of the opponent, when inferring from public information, is governed by Deep Random assumption, that we introduce.

### Shannon and the need of Bayesian inference

Shannon, in [1], established his famous impossibility result, stating that, in order to obtain perfect secrecy in an encryption system, it is needed that the probability of the clear message $P(M)$ is equal to the conditional probability $P(M|E)$ of the clear message knowing the encrypted message. In the case where the encryption system is using a secret key $K$ with a public transform procedure to transform the clear message in the encrypted message, Shannon then comes to the conclusion that perfect secrecy can only be obtained if $H(K) \geq H(M)$ (where $H$ designates Shannon's entropy).

It is a common belief in the cryptologic community that, in cases where the legitimate partners initially shares no secret information (which we can write $H(K) = 0$), the result of Shannon thus means that it is impossible for them to exchange a perfectly secret bit of information. The support for that belief is that, by applying Shannon's method, the absence of key entropy makes the conditional expectation $E[M|E]$ to be completely and equally known by all the parties (legitimate receiver and opponent), as :

$$E[M|E] = \sum_m m \frac{P(E|m)P(m)}{\sum_{m'} P(E|m')P(m')}$$

and thus, that the legitimate receiver cannot gain any advantage over the opponent when he tries to estimate the secret clear message from the public encrypted message. This reasoning however supposes that all the parties have a full knowledge of the distribution $P(M)$, enabling them to perform the above bayesian inference to estimate $M$ from $E$.

*(*) See contact and information about the author at last page*

Shannon himself warned the reader of [1] to that regard, but considered that this assumption is fairly reasonable (let's remember that computers were almost not yet existing when he wrote his article):

*« There are a number of difficult epistemological questions connected with the theory of secrecy, or in fact with any theory which involves questions of probability (particularly a priori probabilities, Bayes' theorem, etc.) when applied to a physical situation. Treated abstractly, probability theory can be put on a rigorous logical basis with the modern measure theory approach. As applied to a physical situation, however, especially when "subjective" probabilities and unrepeatable experiments are concerned, there are many questions of logical validity. For example, in the approach to secrecy made here, **a priori probabilities of various keys and messages are assumed known by the enemy cryptographer.** »*

The model of security that we develop in this article, by enabling the legitimate partners to use a specific form of randomness where the a priori probabilities of the messages cannot be efficiently known by the opponent, puts this opponent in a situation where the above reasoning based on Bayesian inference no longer stands.

**Prior probabilities theory**

The art of prior probabilities consists in assigning probabilities to random event in a context of partial or complete uncertainty regarding the probability distribution governing that random event. The first author who has rigorously considered this question is Laplace [10], proposing the famous *principle of insufficient reason* by which, if an observer do not know the prior probability of occurrence of 2 events, he should consider them as equally likely. In other words, if a random variable $V$ can take several values $v_1, ..., v_n$, and if no information regarding the prior probabilities $P(V = v_i)$ is available for the observer, he should assign them $P(V = v_i) = 1/n$ in any attempt to produce inference from an experiment of $V$.

Several authors observed that this principle can lead to conclusion contradicting the common sense. They proposed some improved principles to assign prior probabilities, like the maximum entropy principle developed by Good [12] and Jaynes [7], or the $c^*$ function of Carnap [11]. The consistency of those improved principles with the logical foundation of probability theory has been carfully and deeply studied.

Jaynes, in [7], has proposed a remarkable and resulted theory to reduce the difficulties associated to the unsifficiently defined Laplace principle. His statement is that, when 2 probability distributions are transformed from one to the other using a finite or isometric transformation group, and if no prior information is available to the observer to privilege one or the other distribution, then, (i) by symmetry, Laplace's principle can be safely applied within the fundamental domain of the group, and (ii) the method matches the maximum entropy criteria. We will also use such symmetry principle in the construction of our protocol.

In all the following, we will only consider measurable sets, where probability can then be intuitively defined over the tribe of all parties of such set.

**Deep random assumption**

We introduce below the Deep Random assumption, that is an objective principle to assign prior probabilities, compatible with the symmetry principle proposed by Jaynes.

Let $V$ be a random variable with values in a set $E$, having unknown (or hidden) probability distributions for $\xi$, except a same public characteristical information $I$. We denote $\Omega_I$ the set of all possible distributions having that same public characteristic information $I$. Let also $U$ be a random variable with values in a set $F$, and whose probability distribution is known and depends from a parameter in set $E$, that parameter being typically the value of an experiment of $V$. If there exists a finite or isometric transformation group $G$ such that, for each transformation $\tau \in G$, the distribution $\Phi_\xi(v)$ has no a priori reason for the opponent $\xi$ to be more likely than $\Phi_\xi(\tau(v))$, (this in particular means that $\Omega_I$ is stable by action of $G$) therefore, from the non-subjectivity principle, the set of possible joint distributions viewed from the opponent, can be restricted to the invariant class by action of $G$:

$$R_G \triangleq \left\{ \frac{1}{|G|} \sum_{\tau \in G} \Phi \circ \tau \mid \forall \Phi \in \Omega_I \right\}$$

If we consider $\varphi$ as any application from $E$ in an Hilbert space $H$, the conditional expectation $E[\varphi(V)|U]_\xi$ can be viewed as the strategy of the opponent $\xi$ to guess private information $\varphi(V)$ from the knowledge of the public information $U$, and the Deep Random assumption then states that the conditional expectation $E[\varphi(V)|U]_\xi$ should be chosen by the opponent within the restricted set of strategies :

$$E[\varphi(V)|U]_\xi \in \Omega_\#(G)$$

$$\triangleq \{\omega(U) = E[\varphi(V)|U] | \Phi(v) \in R_G\} \quad (A)$$

$\Omega_\#(G)$ can in other words be written :

$$\Omega_\#(G) = \left\{ \omega(U) = \frac{\int \varphi(v) P(U|v) \Phi(v) dv}{\int P(U|v) \Phi(v) dv} \mid \Phi(v) \in R_G \right\}$$

**The security model**

We consider Autonomous Entities (AE), which, like in every classical protocol modelization, are entities capable to Generate random bit strings, Publish bit strings on the public channel, Read bit strings published by other AE from the public channel, Store bit strings, Make calculation on bit strings. The main difference of our model is that random generation also includes Deep Random generation, a form of randomness in which the probability distribution used by an AE is made unknown for all other party by the use of a Deep Random Generator (DRG), we will see below how such DRG can be designed.

In our model, an observer (or opponent) AE called $E$ is supposed to have unlimited calculation and storage power, it is also supposed to have full access to the information published by the legitimate AE partners on the public channel, that is supposed to be perfectly accessible and equivalent for any playing party (legitimate partners and opponents). The considered opponent is passive.

The specificity of our model is that when $E$ desires to infer the secret information generated by partner $A$'s DRG (or $B$'s DRG) from a public information, it can only do it in respect of the Deep Random assumption $(A)$ presented above.

This assumption is fairly reasonable, as established by the Prior Probability theory presented above, under the condition that the DRG can actually produce distributions that are undistinguishable and unpredictable among the set $\Omega_I$ of distributions compliant with the public characteristic information $I$. It is also easy to see that for any non-objective strategy to assign prior probabilities, a better objective strategy exists ; « better » being understood under the sense of quadratic optimization in the considered Hilbert space, where $\varphi(V)$ is evaluated.

The protocols that we consider here will be further modelized in the description below, but at this stage one can already say that they obey the Kerckhoffs's principle by the fact that their specifications are entirely public. We can thus modelize the usage of such protocol in 2 phases :

|  | Legitimate partners | Opponent |
| --- | --- | --- |
| The elaboration phase | The specification of the protocol are made public ($\mathcal{P}$ with below notations) | The opponent elaborate its « best » objective strategy, being a deterministic or probabilistic function taking as parameters the public information that are released during an instantiation ($\omega(\cdot,\cdot)$ with below notations) |
| The instantiation phase | The legitimate partners both calculate their estimation of the common secret information based on their part of the secret information generated during an instantiation plus the public information that are released during the instantiation ($V_A(x,j)$ and $V_B(y,i)$ with below notations) | The Opponent calculates its estimation of the common secret information as the value of the strategy function with the released public information as parameters. ($\omega(i,j)$ with below notations) |

Then, when we say that distributions $\Phi_\xi(v)$ and $\Phi_\xi(\tau(v))$ are undistinguishable and unpredictable among the set $\Omega_I$ through Deep Randomness, we mean that they are undistinguishable and unpredictable among the set $\Omega_I$ through Deep Randomness for the opponent at the elaboration phase.

**Concepts of Perfect Secrecy Protocols (PSP)**

The main purpose of the following is to introduce how to design a « Perfect Secrecy Protocol » under Deep Random assumption, and to introduce how to generate Deep Random from classical computing resources.

A « Perfect Secrecy Protocol » (PSP) $\mathcal{P}$, is a protocol according to the general model presented below,

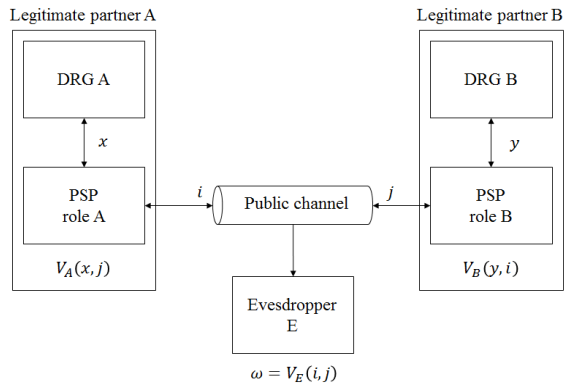

in which the AE $A$ (resp. $B$) requests its Deep Random Generator (DRG) at a given moment to obtain an experiment $x$ (resp. $y$) of a random variable $X$ (resp. $Y$) with hidden probability distribution ; $A$ (resp. $B$) publishes the set of information $i$ (resp. $j$) on the public channel along the protocol ; $A$ (resp. $B$) calculates its estimation of a joined secret information $V_A(x,j)$ (resp. $V_B(y,i)$), with value in an Hilbert space, and the eavesdropping opponent $E$ who has a full access to the public information, calculates its own estimation $V_E(i,j) = \omega(i,j)$, also called strategy of the opponent.

In case of PSP, there must exist a finite or isomorphic group of transformation $G$ keeping $\Omega_I$ stable, such that, for each transformation $\tau \in G$, the joint distribution $\Phi(x,y)$ is, by construction of the Deep Random generation process, undistinguishable from $\Phi(\tau(x,y))$ by the opponent. And therefore, the set of possible joint distributions viewed from the opponent, can be restricted to the invariant class by action of $G$ :

$$R_G \triangleq \left\{ \frac{1}{|G|} \sum_{\tau \in G} \Phi \circ \tau(x,y) \,|\, \forall \Phi \in \Omega_I \right\}$$

From the Deep Random assumption, the set of optimal strategies for the opponent can be restricted to :

$$\Omega_\#(G, \mathcal{P}) \triangleq \{\omega(i,j) = E[V_A|i,j] | \Phi(x,y) \in R_G\}$$

The set of all possible such groups $G$ for the considered protocol $\mathcal{P}$, is denoted $\Gamma(\mathcal{P})$.

Then, the protocol $\mathcal{P}$ is a PSP if it verifies the following property $(P)$ :

$$\exists G \in \Gamma(\mathcal{P}), \alpha > 1 | \forall \omega \in \Omega_\#(G, \mathcal{P}): E[\|\omega - V_A\|^2] \geq \alpha E[\|V_B - V_A\|^2] \quad (P)$$

In particular, this implies that there exists, for any strategy $\omega$ (even outside of $\Omega_\#$) chosen by the opponent, a couple of distributions $\Psi$ and $\Psi'$ such that, if they are used respectively by $A$ and $B$ to generate $x$ and $y$ :

$$E[\|\omega - V_A\|^2]_{\Psi,\Psi'} \geq \alpha E[\|V_B - V_A\|^2]_{\Psi,\Psi'} \quad (P')$$

■ Indeed, $(P)$ shall in particular be true for an element of $R_G$ corresponding to $\Phi \mapsto \widetilde{\Phi} = \frac{1}{|G|} \sum_{\tau \in G} \Phi \circ \tau(x,y)$ where $\Phi$ represents the actual joined distribution used by the legitimate partners. Then, by considering the obvious :

$$E[\|\omega - V_A\|^2]_{\widetilde{\Phi}} \geq \inf_{\omega'} E[\|\omega' - V_A\|^2]_{\widetilde{\Phi}}$$

the inf in the above inequality is reached for an element $\omega^*$ of $\Omega_\#(G, \mathcal{P})$. Thus, from $(P)$ :

$$E[\|\omega - V_A\|^2]_{\widetilde{\Phi}} \geq E[\|\omega^* - V_A\|^2]_{\widetilde{\Phi}} \geq \alpha E[\|V_B - V_A\|^2]_{\widetilde{\Phi}}$$

The last inequality being also possibly written :

$$\int_{x,y,i,j} \|\omega^* - V_A\|^2 P(i,j|x,y) \widetilde{\Phi}(x,y) d_{x,y}$$
$$\geq \alpha \int_{x,y,i,j} \|V_B - V_A\|^2 P(i,j|x,y) \widetilde{\Phi}(x,y) d_{x,y}$$

The above implies that there exits $(x,y)$ such that :

$$\int_{i,j} \|\omega^* - V_A\|^2 P(i,j|x,y) \geq \alpha \int_{i,j} \|V_B - V_A\|^2 P(i,j|x,y)$$

Which gives the result for $\Psi = \delta_x$ and $\Psi' = \delta_y$. ■

An example of PSP is given in Annex. Its detailed presentation, and proof of secrecy under Deep Random assumption can be found in [9] Section III.

In practice, one can obtain the property $(P)$ typically by deriving the information $i$ (resp. $j$) published by $A$ (resp. $B$) from a degradation of a secret information $x$ (resp. $y$), in a way that the only mean for the opponent to estimate say $V_A(x,j)$ from the public information $\{i,j\}$ is to perform a bayesian inference from $\{i,j\}$, which requires the knowledge of the probability distribution of $x$.

Degradation can be understood as degradation of the accuracy of the signal $X$.

Let's define rigorously what a Degradation is. Let $V$ be a random variable with values in a set $E$ and probability distribution $\Phi$ ; let also $U$ be be a random variable with values in a set $F$ whose probability distribution depends on a parameter in set $E$, and that is drawn after an experiment of $V$ giving the input parameter. Let $H$ be a Hilbert space, and $\varphi : E \mapsto H$ be an application called « evaluation ».

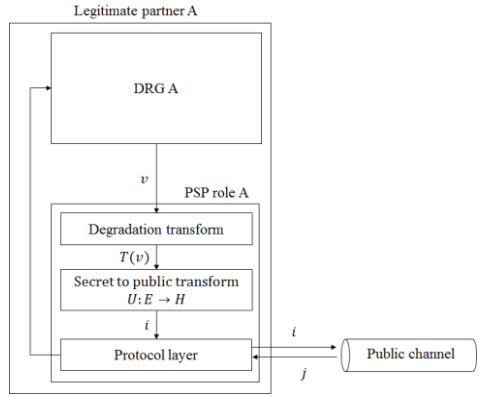

**Definition :**

*A Degradation in $E$ related to $V, U$ and $\varphi$ is an application $T : E \mapsto E$, transforming the output of $V$, and such that, for any applications $\omega_T : F \mapsto H$, verifying $E[\omega_T(U)|V] = \varphi(V)$, the following strict inequality stands :*

$$E[\|\omega_T(U) - \varphi(V)\|^2] > \inf_\omega E[\|\omega(U) - \varphi(V)\|^2]$$

In this definition, the notations should be read, due to transformation $T$, as :

$$E[\omega_T(U)|V] = \int_u \omega_T(u) P(u|T(V))$$

$$E[\|\omega_T(U) - \varphi(V)\|^2] = \int_v \int_u \|\omega_T(u) - \varphi(v)\|^2 P(u|T(v)) \, \Phi(v) dv$$

A Degradation can be seen as a transform that is applied on a secret information, in order to reduce the accuracy of the derived public information. As an analogy with quantum behavior (Heisenberg uncertainty principle), the above definition means that, with Degradation, it is impossible to infer the private information from the public information with equaling both the first and second moments.

Let's give an example that will be used in the protocol that is presented in Annex as a PSP example. $E = [0,1]^n \times [0,1]^n$, $H = \mathbb{R}$, and $U$ is a pair of Bernoulli vector random variables $(i,j)$ with values in $\{0,1\}^n \times \{0,1\}^n$, and whose parameter vectors are in $E$ and coming from an expriment of $V$. The Degradation is given, for any $V = (x = (x_1, \ldots, x_n), y = (y_1, \ldots, y_n)) \in E$, by :

$$T_k((x,y)) = \left(\left(\frac{x_1}{k}, \ldots, \frac{x_n}{k}\right), \left(\frac{y_1}{k}, \ldots, \frac{y_n}{k}\right)\right)$$

where $k > 1$. The application $\varphi$ is chosen as :

$$\varphi(x,y) = \frac{x.y}{n}$$

(where $x.y$ designates the scalar product of $x$ and $y$). It is easy to show (you can easily verify by yourself) that the only possible applications $\omega_T$ is :

$$\omega_T(i,j) = k^2 \frac{i.j}{n}$$

and that :

$$E\left[\left(k^2 \frac{i.j}{n} - \frac{x.y}{n}\right)^2\right] \geq (k - O(1)) \inf_\omega E\left[\left(\omega_{i,j} - \frac{x.y}{n}\right)^2\right]$$

Which means that $T$ is a Degradation for $k$ sufficiently large.

**Deep random generation**

It is of course of great importance to be able to concretely produce some Deep Random if we want to use it for practical applications like cryptography. A Deep Random Generator (DRG) must be able to produce a random signal for which a large diversity of distributions could be considered as equally likely for the observer. By using an objective method for assigning prior probabilities, the observer has to choose a strategy prior to the release of the public information. This strategy is an application $\omega : F \mapsto H$, where $F$ is the set of possible values for the public information, and $H$ is the Hilbert space in which the joined secret information is projected. And by doing so, the DRG should then ensure that any set of possible distributions $\{\Phi_1, \ldots, \Phi_S\}$ should remain indistinguishable by any such prior strategy. This enables us to give a definition for the concept of indistinguishable distributions (with the same notations than previously) :

**Definition 2 :**

*A set of distributions $\{\Phi_1, \ldots, \Phi_S\}$ is said made of $\alpha-$undistinguishable distributions $(\alpha > 1)$ relatively to $Y$ and $\varphi(X)$ if :*

$$\inf_\omega \frac{1}{S}\sum_s \int \|\omega(y) - \varphi(x)\|^2 P(y|x)\Phi_s(x)dx$$
$$> \alpha \frac{1}{S}\sum_s \inf_{\omega_s|\Phi_s} \int \|\omega_s(y) - \varphi(x)\|^2 P(y|x)\Phi_s(x)dx$$

Let's explain the relevance of this definition. If the distributions were not undistinguishable relatively to $Y$ and $\varphi(X)$, there would exist a strategy $\omega$ enabling the observer to determine from an experiment of $Y$ what distribution among $\{\Phi_1, \ldots, \Phi_S\}$ is the most likely to be used, with the same level of accuracy to estimate $\varphi(X)$ than with having knowledge of which $\Phi_s$ is actually used. And from there, the observer could do a classical bayesian inference of $\varphi(x)$ with the determined $\Phi_s$, which would contradict the inequality above.

It may appear impossible at first sight to generate an unknowledgeable probability distribution from a deterministic computing resource. In fact it is not. The basic idea is to use infinite incrementing counters as primary source of entropy, and then to run across those counters one or several Cantor' style diagonal recursive constructing process, in which at each step $m + 1$, the generator picks (through classical randomness) a new distribution that defeats the best possible strategy knowing all the past distributions for $t \leq m$. Such a DRG, based on continuous and recursive constructing process, can only be designed in association with a given PSP. At each step $m + 1$, the generator emulates the PSP internally and picks (through classical randomness) a new couple of distributions that defeats the best possible strategy knowing all the past distributions for $t \leq m$, which is always possible for a PSP as as per property $(P')$ shown above. The source of secret entropy is the current values of the inifinite counters (several can run in parallel) of the continuous recursive process, together with the classical random that is used at each step to pick a defeating distribution. Such entropy must of course be at least equal to the length of the message to send to obtain perfect secrecy.

A detailed presentation of a DRG using a recursive method as introduced above, associated with the example of PSP, can be found in [9] Section IV.

**Conclusion**

We have introduced a new idea to design secure communication protocol able to reach perfect secrecy. The obtained result, apparently contradicting Shannon's pessimistic theorem, is possible thanks to the nature of Deep Random that prevents the use of Bayesian inference from public information, because not only draws but also probability distributions themselves are unknown to the opponent. In this work, our main objective was to expose this new idea and to present a working protocol (Annex). But of course, once done, the next question is : if such protocol exist, what is the best one ? In cryptology and communication sciences, there exist many criteria of quality for a protocol. One of the most critical one for practical implementation is the burden of exchanged data needed to obtain a secure digit (or plain text) of information. The question then roughly becomes: given $\varepsilon, \varepsilon' > 0$ (introduced as error rate and opponent knowledge rate in Annex), what is, under Deep Random assumption, a protocol exchanging the minimal quantity to obtain the above upper-bounds ?

We introduce this more general question in [9], and we hope this challenge will create enthusiasm for the largest possible number of curious minds interested in cryptology, information theory and communication sciences.

**Who is the author ?**
I have been an engineer in computer science for 20 years. My professional activities in private sector are related to IT Security and digital trust, but have no relation with my personal research activity in the domain of cryptology. If you are interested in the topics introduced in this article, please feel free to establish first contact at tdevalroger@gmail.com

**Annex: example of a Perfect Secrecy Protocol.**

In this Annex we present an example of PSP, based on Bernouilli random variables. The detailed presentation and the proof of secrecy under Deep Random assumption can be found in [9].

For this example, besides being hidden to the opponent, the probability distribution used by each legitimate partner also needs to have specific properties in order to prevent the opponent to evaluate $V_A$ with the same accuracy than the legitimate partners by using symmetry and independence in the choice of their distributions by the legitimate partners.

Those specific properties are :

(i) Each probability distribution $\Phi$ (of $A$ and $B$) must be « far » from its symmetric projection $\bar{\Phi}(x) = \frac{1}{n!}\sum_{\sigma \in \mathfrak{S}_n} \Phi \circ \sigma(x)$ (where $\sigma(x)$ represents $(x_{\sigma(1)}, \ldots, x_{\sigma(n)})$)
(ii) At least one of the distribution (of $A$ or $B$) must avoid to have brutal variations (Dirac)

The technical details explaining those constraints are presented in [9]. The set of compliant distributions is denoted $\zeta(\alpha)$ where $\alpha$ is a parameter that measures the « remoteness » of a distribution from its symmetric projection.

For such a distribution $\Phi$, a tidying permutation, denoted $\sigma_\Phi$, is a specific permutation that enables to give a canonical form $\Phi \circ \sigma_\Phi$ of $\Phi$, such form being useful to synchronize two distributions by transitivity. Again, technical details are given in [9].

Here are the steps of the proposed protocol :

$A$ and $B$ are two AE, called the legitimate partners. The steps of the protocol $\mathcal{P}(\alpha, n, k)$ are the followings :

*Step 1 – Deep Random Generation : A and B pick independantly the respective probability distributions $\Phi$ and $\Phi' \in \zeta(\alpha)$, so that $\Phi$ (resp. $\Phi'$) is secret (under Deep Random assumption) for any observer other than A (resp. B) beholding all the published information. A draws the parameter vector $x \in [0,1]^n$ from $\Phi$. B draws the parameter vector $y \in [0,1]^n$ from $\Phi'$.*

*Step 2 – Degradation : A generates a Bernouilli experiment vectors $i \in \{0,1\}^n$ from the parameter vector $\frac{x}{k}$. A publishes i. B generates a Bernouilli experiment vectors $j \in \{0,1\}^n$ from the parameter vector $\frac{y}{k}$. B publishes j.*

*Step 3 – Dispersion : A and B also pick respectively a second probability distribution $\Psi$ and $\Psi' \in \zeta(\alpha)$ such that it is also secret (under Deep Random assumption) for any observer other than A (resp. B). $\Psi$ is selected also such that $\int_{|x| \in [k|i|-\sqrt{n}, k|i|+\sqrt{n}]} \Psi(x)dx \geq \frac{1}{2\sqrt{n}}$ in order to ensure that $|i|$ is not an unlikely value for $\approx \left|\frac{x}{k}\right|$ (same for $\Psi'$ by replacing x by y and i by j). $\Psi$ (resp. $\Psi'$) is used to scramble the publication of the tidying permutation of A (resp. B). A (resp. B) calculates a permutation $\sigma_d[i]$ (resp. $\sigma'_d[j]$) representing the reverse of the most likely tidying permutation on $\Psi$ (resp. $\Psi'$) to produce i (resp. j). In other words, with i, $\sigma_d[i]$ realizes :*

$$\max_{\sigma \in \mathfrak{S}_n} \int_x P(i|x)\Psi \circ \sigma_\Psi \circ \sigma^{-1}(x)dx$$

*Then A (resp. B) draws a boolean $b \in \{0,1\}$ (resp. $b'$) and publishes in a random order $(\mu_1, \mu_2) = t^b(\sigma_d[i], \sigma_\Phi)$, (resp. $(\mu'_1, \mu'_2) = t^{b'}(\sigma'_d[j], \sigma_{\Phi'})$) where t represents the transposition of elements in a couple.*

*Step 4 – Synchronization : A (resp. B) chooses randomly $\sigma_A$ (resp. $\sigma_B$) among $(\mu'_1, \mu'_2)$ (resp. $(\mu_1, \mu_2)$).*

*Step 5 – Decorrelation: A computes $V_A = \frac{\sigma_\Phi^{-1}(x).\sigma_A^{-1}(j)}{n}$, B computes $V_B = \frac{\sigma_B^{-1}(i).\sigma_{\Phi'}^{-1}(y)}{n}$. $V_A$ and $V_B$ are then transformed respectively by A and B in binary output thanks to a sampling method described hereafter. At this stage the protocol can then be seen as a broadcast model with 2 Binary Symmetric Channels (BSC), one between A and B and one between A and $\xi$ who computes a certain $V_\xi$, called $\xi$'s strategy, that is to be transformed in binary output by the same sampling method than for A and B. It is shown in Theorem 1 that those 2 BSC are partially independent, which enable to create Advantage Distillation as shown in [5].*

*Step 5' – Advantage Distillation : by applying error correcting techniques with code words of length L between A and B, as introduced in [5], we show in Theorem 1 that we can then create advantage for B compared to $\xi$ in the binary flows resulting from the error correcting code.*

*Step 6 : classical Information Reconciliation and Privacy Amplification (IRPA) techniques then lead to get accuracy as close as desired from perfection between*

*estimations of legitimate partners, and knowledge as close as desired from zero by any unlimitedly powered opponent, as shown in [4].*

The choice of the parameters $(\alpha, n, k, L)$ are set to make steps 5, 5' and 6 possible, details are given in [9].

The Degradation transformations $x \mapsto \frac{x}{k}$ and $y \mapsto \frac{y}{k}$ with $k > 1$ at step 2 are the ones that prevent the use of direct inference by the opponent, and of course, the Deep Random Generation at step 1 prevents the use of Bayesian inference based on the knowledge of the probability distribution. The synchronization step 4 is designed to overcome the independence between the choices of the distributions of $A$ and $B$, and needs that the distributions to have special properties ($\in \zeta(\alpha)$) in order to efficiently play their role. It is efficient in 1/4 of cases (when $B$ picks $\sigma_B = \sigma_\Phi$ and $A$ picks $\sigma_A = \sigma_{\Phi'}$, which we will call favorable cases). And to prevent $\xi$ from gaining knowledge of $\sigma_\Phi$, Dispersion step 3 mixes $\sigma_\Phi$ within $(\mu_1, \mu_2)$ with another permutation $\sigma_d[i]$ (and $\sigma_{\Phi'}$ within $(\mu'_1, \mu'_2)$ with another permutation $\sigma'_d[j]$) that (1) is undistinguishable from $\sigma_\Phi$ knowing $i$, and (2) manages to make the estimation of $\xi$ unefficient as shown in [9]. We denote the following set of strategies (invariant by transposition of $(\mu_1, \mu_2)$ or $(\mu'_1, \mu'_2)$):

$$\Omega'_\# \triangleq \left\{ \omega(i, j, (\mu_1, \mu_2), (\mu'_1, \mu'_2)) \,\middle|\, \forall b, b' \in \{0,1\}: \omega\left(i, j, \tau^b(\mu_1, \mu_2), \tau^{b'}(\mu'_1, \mu'_2)\right) = \omega(i, j, (\mu_1, \mu_2), (\mu'_1, \mu'_2)) \right\}$$

Because of Deep Random assumption ($A$) over the group $\{Id, t\}$ applied to the distribution of $(\mu_1, \mu_2)$ and $(\mu'_1, \mu'_2)$, the strategy of the opponent can thus be restricted to $V_\xi \in \Omega'_\#$.

$i$ is entirely determined by $|i|$ and a permutation, which explains the constraint and transformation applied on $\Psi$ in step 3 to make $\sigma_\Phi$ and $\sigma_d[i]$ indisguishable knowing $i$ (same with $\sigma'_d[j]$, $\sigma_{\Phi'}$, and $j$).

The synchronization step has a cost when considering the favorable cases: $\xi$ knows that $\Phi$ and $\Phi'$ are synchronized in favorable cases, which means in other words that $\xi$ knows that an optimal (or quasi optimal) permutation is applied to $\Phi'$. This also means that in favorable cases, all happen like if when A picks $\Phi \circ \sigma$ instead of $\Phi$, the result of the synchronization is that B uses $\Phi' \circ \sigma$ instead of $\Phi'$. Starting from the most general strategy $\omega \in \Omega'_\#$ for $\xi$, we will also consider in the proof of main Theorem the following additional restrictions applicable to the favorable cases:

- Restriction to the strategies of the form $\omega(i, j)$, because $(\sigma_d[i], \sigma'_d[j])$ depends only on $(i, j)$ and not on $\Phi$ neither $\Phi'$,
- And then restriction to the set of strategies such that $\omega_{i,j} = \omega_{\sigma(i), \sigma(j)}, \forall \sigma \in \mathfrak{S}_n$, in other words strategies invariant by common permutation on $i, j$.

which leads to define the more restricted set of strategies:

$$\Omega_\#(G, \mathcal{P}) = \left\{ \omega \in [0,1]^{2^{2n}} \,\middle|\, \omega(\sigma(i), \sigma(j), \mu_1, \mu_2, \mu'_1, \mu'_2) = \omega(\sigma(i), \sigma(j)) = \omega(i, j) \,, \forall \sigma \in \mathfrak{S}_n \right\}$$

Regarding the legitimate partners, when $B$ picks $\sigma_B = \sigma_\Phi$ and $A$ picks $\sigma_A = \sigma_{\Phi'}$, the choice of $\sigma_A$ and $\sigma_B$ remain independant from $i, j$, so that $i$ and $j$ remain draws of independant Bernouilli random variables, then allowing to apply Chernoff-style bounds for the legitimate partners. When $B$ picks $\sigma_B = \sigma_d[i]$ or $A$ picks $\sigma_A = \sigma'_d[j]$, this is no longer true and $V_B$ or $V_A$ become erratic, which will lead to error detection by error correcting code at step 5'.

The heuristic table analysis of the protocol is then the following:

|  | $B$ picks $\sigma_B = \sigma_\Phi$ among $(\mu_1, \mu_2)$ | $B$ picks $\sigma_B = \sigma_d[i]$ among $(\mu_1, \mu_2)$ |
|---|---|---|
| $A$ picks $\sigma_A = \sigma_{\Phi'}$ among $(\mu'_1, \mu'_2)$ | $A$ and $B$ respective estimations are close in ~100% of cases, and thus both obtain accurate estimation of the combined shared secret in ~100% of cases. $\xi$ cannot make accurate estimation of the combined shared secret in at least ~25% of cases (if $\xi$ tries to have a strategy depending on $(\mu_1, \mu_2, \mu'_1, \mu'_2)$, then $(\sigma_d[i], \sigma'_d[j])$ is indistinguishable from $(\sigma_\Phi, \sigma_{\Phi'})$ and is thus picked by $\xi$ in 25% of cases. | $A$ and $B$ respective estimations are not close which leads to error detection and finally discarding. |
| $A$ picks $\sigma_A = \sigma'_d[j]$ among $(\mu'_1, \mu'_2)$ | $A$ and $B$ respective estimations are not close which leads to error detection and finally discarding. | $A$ and $B$ respective estimations are not close which leads to error detection and finally discarding. |

This is a heuristic reasoning, and we must rather consider most general strategies $\omega(i,j,\mu_1,\mu_2,\mu'_1,\mu'_2)$ and write the probability equations with the appropriate group transform, under Deep Random assumption. But this little array explains why we create partial independence between the BSC and consequently then an advantage for the legitimate partners compared to the opponent, bearing in mind that $(\mu_1,\mu_2)$ (resp. $(\mu'_1,\mu'_2)$) are absolutely undistinguishable knowing $i$ (resp. $j$), due to the fact that the distributions $\Phi$ and $\Psi$ (resp. $\Phi'$ and $\Psi'$) are unknown and thus also absolutely undistinguishable by $\xi$.